\documentclass[aps,prl,preprint,amsmath,amssymb]{revtex4-1}
\usepackage{graphicx}

\begin{document}

\title{Limiting fluctuations in quantum gravity to diffeomorphisms}

\author{Brian Slovick}
\altaffiliation{brian.slovick@sri.com}
\affiliation{Applied Optics Laboratory, SRI International, Menlo Park, California 94025, USA}

\date{\today}

\begin{abstract}
Within the background field formalism of quantum gravity, I show that if the quantum fluctuations are limited to diffeomorphic gauge transformations rather than the physical degrees of freedom, as in conventional quantum field theory, all the quantum corrections vanish on shell and the effective action is equivalent to the classical action. In principle, the resulting theory is finite and unitary, and requires no renormalization. I also show that this is the unique parameterization that renders the path integral independent of the on-shell condition for the background field, a form of background independence. Thus, a connection is established between background independence and renormalizability and unitarity.
\end{abstract}

\maketitle

\section{Introduction}
The development of a quantum field theory of gravitation remains a challenge. Since the gravitational coupling constant has units of length, the quantum corrections correspond to higher-derivative terms with divergent coefficients \cite{tHooft1974a,Alvarez1989,Hamber2009}. At one-loop order, the divergence in pure gravity vanishes on shell, i.e., when the classical equations of motion are imposed on the background fields \cite{tHooft1974a,Alvarez1989,Kallosh1978,tHooft1974b,Hamber2009,Kallosh1978,Capper1983,Slovick2018}. However, on-shell divergences appear at two-loop order \cite{Goroff1985,Goroff1986,Ven1992,Bern2017} or when coupled to matter \cite{tHooft1974a,Deser1974}. In order to renormalize these terms, counterterms of a similar form must be included in the bare classical action \cite{Stelle1977,Tomboulis1980,Fradkin1982,Asorey1997}. However, the addition of higher derivative terms leads to ghosts and a violation of unitarity in flat space perturbation theory \cite{Stelle1977,Tomboulis1980,Shapiro2015}. Therefore, approaches to eliminate on-shell divergences are needed to obtain a unitary and renormalizable theory.

The most common approach for calculating the divergences in quantum gravity is to obtain the effective action using the background field method \cite{Vilkovisky1984,Barvinsky1985,Huggins1988,Buchbinder1992,Avramidi1995,Goncalves2018,Lavrov2019,Giacchini2019,Giacchini2020,Giacchini2020b}. In general, the effective action is given by a functional integral of the classical action over quantum field configurations. In the background field method, the quantum field is divided into a sum of a fixed classical background field and a quantum fluctuation, and the action is expanded perturbatively around the background field \cite{Goncalves2018,Lavrov2019,Giacchini2019,Giacchini2020}. The action is then supplemented with gauge-fixing and ghost terms and the functional integral is performed over the gauge-fixed quantum fluctuation \cite{Goncalves2018,Lavrov2019,Giacchini2019,Giacchini2020,Giacchini2020b}. In a renormalizable quantum field theory, this procedure generates divergences of a similar form to the classical action, allowing them to be expressed as renormalizations of the mass, field, and coupling constants \cite{Hamber2009,Peskin2018}. In gravity, due to the dimensionality of the coupling constant, the divergences correspond to higher order curvature invarianta, which are not renormalizable and violate unitarity.

In this Letter, I show that if the quantum fluctuations of the background field are constrained to diffeomorphic gauge transformations rather than the physical degrees of freedom, as in conventional quantum field theory, all of the quantum corrections vanish on shell, and the effective action is equal to the classical action with no quantum corrections. In principle, the resulting theory is unitary and requires no renormalization. I also show that this choice of fields renders the effective action independent of the on shell condition for the background fields, and thus incorporates a form of background independence. Lastly, I discuss the implications of associating quantum fluctuations with gauge transformations and the relation to other renormalizable field theories.

\section{Approach}
The classical Einstein-Hilbert action for the gravitational field $g_{\mu\nu}$ is
\begin{equation}
S[g_{\mu\nu}]=\frac{1}{16\pi G} \int d^4x \sqrt{-g} g^{\mu\nu}R_{\mu\nu},
\end{equation}
where $G$ is Newton's constant and $R_{\mu\nu}$ is the Ricci tensor. The corresponding quantum theory is described by the effective action $\Gamma[\bar{g}_{\mu\nu}]$, which is a functional of the classical mean field $\bar{g}_{\mu\nu}$. In the background field method, the effective action is obtained from a functional integral of the form \cite{Vilkovisky1984,Barvinsky1985,Huggins1988,Buchbinder1992,Avramidi1995,Goncalves2018,Lavrov2019,Giacchini2019,Giacchini2020,Giacchini2020b}
\begin{eqnarray}
\exp \left( \frac{i}{\hbar} \Gamma[\bar{g}_{\mu\nu}] \right)=\int \mathcal{D}h_{\mu\nu} \exp \left( \frac{i}{\hbar} S[\bar{g}_{\mu\nu}+h_{\mu\nu}] -\frac{i}{\hbar}\int d^4x \sqrt{-\bar{g}}h^{\mu\nu}\frac{\delta \Gamma[\bar{g}_{\mu\nu}]}{\delta \bar{g}^{\mu\nu}}\right),
\end{eqnarray}
where $g_{\mu\nu}$ has been divided into the sum of $\bar{g}_{\mu\nu}$ and the quantum fluctuation $h_{\mu\nu}$ as \cite{Abbott1981,Hamber2009,Giacchini2019}
\begin{equation}
g_{\mu\nu}=\bar{g}_{\mu\nu}+h_{\mu\nu}.
\end{equation}
At this stage, one normally proceeds by expanding $S[\bar{g}_{\mu\nu}+h_{\mu\nu}]$ around $\bar{g}_{\mu\nu}$, adding gauge fixing and ghost terms, and performing the functional integral over $h_{\mu\nu}$ \cite{Vilkovisky1984,Barvinsky1985,Huggins1988,Buchbinder1992,Avramidi1995,Goncalves2018,Lavrov2019}. However, this procedure leads to nonrenormalizable and nonunitary divergences on shell. This can be seen by recognizing Eq. (2) as the functional Fourier transform of $\exp \left( \frac{i}{\hbar} S[\bar{g}_{\mu\nu}+h_{\mu\nu}]\right)$ with respect to the conjugate variables $h_{\mu\nu}$ and $\delta \Gamma[\bar{g}_{\mu\nu}]/\delta \bar{g}^{\mu\nu}$. If $\exp \left( \frac{i}{\hbar} S[\bar{g}_{\mu\nu}+h_{\mu\nu}]\right)$ is independent of $h_{\mu\nu}$, it can be factorized and the functional integral over $h_{\mu\nu}$ reduces to a delta functional in the on shell condition $\delta \left[ \delta \Gamma[\bar{g}_{\mu\nu}]/\delta \bar{g}^{\mu\nu} \right]$. Alternatively, any dependence of $S[\bar{g}_{\mu\nu}+h_{\mu\nu}]$ on $h_{\mu\nu}$ will break the delta functional form, and thus produce nonrenormalizable terms on shell. Therefore, the only way to ensure that all quantum corrections vanish on shell, and thus preserve unitarity, is for $S[\bar{g}_{\mu\nu}+h_{\mu\nu}]$ to be independent of $h_{\mu\nu}$, that is $h_{\mu\nu}$ must correspond to a diffeomorphic gauge transformation
\begin{equation}
h_{\mu\nu}=\nabla_\mu \xi_\nu+\nabla_\nu \xi_\mu.
\end{equation}
In this case, $S[\bar{g}_{\mu\nu}+h_{\mu\nu}]=S[\bar{g}_{\mu\nu}]$ and the expression for the effective action reduces to
\begin{eqnarray}
\exp \left( \frac{i}{\hbar} \Gamma[\bar{g}_{\mu\nu}] \right)=\exp \left( \frac{i}{\hbar} S[\bar{g}_{\mu\nu}] \right)  \int \mathcal{D}\xi_\rho  \Delta[h_{\mu\nu}] \exp \left( -\frac{i}{\hbar}\int d^4x \sqrt{-\bar{g}} \nabla^{(\mu} \xi^{\nu)} \frac{\delta \Gamma[\bar{g}_{\mu\nu}]}{\delta \bar{g}^{\mu\nu}}\right),\nonumber
\end{eqnarray}
where $\Delta[h_{\mu\nu}]$ is the Jacobian determinant of the diffeomorphism \cite{Green1988,Polchinski2005,Zamolodchikov2007,Erbin2015,Erbin2021} 
\begin{equation}
\Delta[h_{\mu\nu}]=\det  \left( \frac{\delta h_{\mu\nu}(x)}{\delta \xi_\rho(x')}\right) = \det \left( \delta_{(\mu}{}^\rho\nabla_{\nu)}\delta(x-x') \right).
\end{equation}
Since the determinant is independent of $\xi_{\rho}$, it can be brought outside the integral. Then, performing integration by parts on the integrand and ignoring the boundary term,
\begin{eqnarray}
\exp \left( \frac{i}{\hbar} \Gamma[\bar{g}_{\mu\nu}] \right)=\exp \left( \frac{i}{\hbar} S[\bar{g}_{\mu\nu}] \right) \Delta[h_{\mu\nu}] \int \mathcal{D}\xi_\rho   \exp \left( \frac{i}{\hbar}\int d^4x \sqrt{-\bar{g}}  \xi^{(\mu} \nabla^{\nu)} \frac{\delta \Gamma[\bar{g}_{\mu\nu}]}{\delta \bar{g}^{\mu\nu}}\right).
\end{eqnarray}
Recognizing the integral over $\xi_\rho$ as a functional delta function, this reduces to
\begin{eqnarray}
\exp \left( \frac{i}{\hbar} \Gamma[\bar{g}_{\mu\nu}] \right)=\exp \left( \frac{i}{\hbar} S[\bar{g}_{\mu\nu}] \right)\Delta[h_{\mu\nu}] \delta \left[\delta^{(\mu}{}_\rho\nabla^{\nu)}\frac{\delta \Gamma[\bar{g}_{\mu\nu}]}{\delta \bar{g}^{\mu\nu}}\right].
\end{eqnarray}
Applying the scaling property for delta functions,
\begin{equation}
\delta \left[\delta^{(\mu}{}_\rho\nabla^{\nu)}\frac{\delta \Gamma[\bar{g}_{\mu\nu}]}{\delta \bar{g}^{\mu\nu}}\right]=\Delta^{-1}[h_{\mu\nu}] \delta \left[\frac{\delta \Gamma[\bar{g}_{\mu\nu}]}{\delta \bar{g}^{\mu\nu}}\right],
\end{equation}
the expression for the effective action reduces to
\begin{eqnarray}
\exp \left( \frac{i}{\hbar} \Gamma[\bar{g}_{\mu\nu}] \right)=\exp \left( \frac{i}{\hbar} S[\bar{g}_{\mu\nu}] \right) \delta \left[ \frac{\delta \Gamma[\bar{g}_{\mu\nu}]}{\delta \bar{g}^{\mu\nu}} \right].
\end{eqnarray}
Therefore, on shell the effective action is equal to the classical action,
\begin{eqnarray}
\frac{\delta \Gamma[\bar{g}_{\mu\nu}]}{\delta \bar{g}^{\mu\nu}}=0, \quad \quad \Gamma[\bar{g}_{\mu\nu}] =S[\bar{g}_{\mu\nu}].
\end{eqnarray}
This is the main result: when the quantum fluctuations are constrained to diffeomorphic gauge transformations, all the quantum corrections vanish on shell. In principle, the resulting theory is finite and unitary and requires no renormalization.

A unique aspect of this approach is that it incorporates a form of background independence, namely the effective action is independent of the on shell condition for the background fields.  In classical general relativity, background independence follows from diffeomorphism invariance and results in a lack of dependence on a fixed, non-dynamical background metric \cite{Rovelli2000,Rovelli2004}. To see how background independence arises in the current approach, consider the variation of the action 
\begin{equation}
S[\bar{g}_{\mu\nu}+h_{\mu\nu}]=S[\bar{g}_{\mu\nu}]-\frac{1}{16\pi G}\int d^4x \sqrt{-\bar{g}} \bar{G}^{\mu\nu}\nabla_{(\mu} \xi_{\nu)},
\end{equation}
where $\bar{G}^{\mu\nu}=\bar{R}^{\mu\nu}-\frac{1}{2}\bar{R}\bar{g}^{\mu\nu}$ is the on-shell condition for the background field. Integrating by parts, the variation can be written as a boundary term, which vanishes under reparameterization, and a term proportional to $\nabla_\mu \bar{G}^{\mu\nu}$, which is zero by the Bianchi identity. Therefore, when $h_{\mu\nu}$ is a diffeomorphism, $S[\bar{g}_{\mu\nu}+h_{\mu\nu}]$, and thus $\Gamma[\bar{g}_{\mu\nu}]$, is independent of the on shell condition for the background fields, which is a form of background independence. 

From the equation of motion for the effective action $\delta \Gamma[\bar{g}_{\mu\nu}]/\delta \bar{g}^{\mu\nu}=-\frac{1}{2}T_{\mu\nu}$, it follows that Eq. (10) applies to pure gravity with no external sources. When sources are present, the more general form involving the Jacobian determinant [Eq. (7)] must be used. This determinant also arises in gauge fixing of conformal field theories. It can be exponentiated using the Faddeev-Popov determinant \cite{Green1988,Polchinski2005,Zamolodchikov2007,Erbin2015,Erbin2021} 
\begin{equation}
\det\left( \frac{\delta h_{\mu\nu}}{\delta \xi_\rho}\right)=\int \mathcal{D}b_{\alpha\nu}\mathcal{D}c^\alpha  \exp \left( i\int d^4x \sqrt{-\bar{g}} \bar{g}^{\mu\nu} c^\alpha   \nabla_\mu b_{\alpha\nu} \right),
\end{equation}
where $c^\alpha$ and $b_{\alpha\nu}$ are anti-commuting ghost fields, and $b_{\alpha\nu}$ is symmetric and traceless. Inserting this expression into Eq. (7), and assuming a conserved stress tensor ($\nabla^\nu T_{\mu\nu}=0$),
\begin{equation}
\exp \left( \frac{i}{\hbar} \Gamma[\bar{g}_{\mu\nu}] \right)=\exp \left( \frac{i}{\hbar} S[\bar{g}_{\mu\nu}] \right) \int \mathcal{D}b_{\alpha\nu}\mathcal{D}c^\alpha \exp \left( i\int d^4x \sqrt{-\bar{g}} \bar{g}^{\mu\nu} c^\alpha   \nabla_\mu b_{\alpha\nu} \right).
\end{equation}
Thus, for gravity coupled to matter with a conserved stress tensor, the effective action is equal to the classical action with additional reparameterization ghosts. These ghosts represent the residual degrees of freedom between $h_{\mu\nu}$ and the vector gauge transformation, and are necessary to maintain consistency of the path integral measure. The energy momentum tensor associated with the ghosts is \cite{Qualls2015,Erbin2021}
\begin{equation}
T_{\mu\nu}=c^\alpha\nabla_\mu b_{\alpha\nu}-\frac{1}{2}g_{\mu\nu}c^\alpha \nabla^\beta b_{\alpha\beta},
\end{equation}
which in flat space perturbation theory would give rise to an additional ghost-graviton vertex linear in the incoming momenta \cite{tHooft1973b}. The equations of motion for the ghosts are \cite{Zamolodchikov2007,Erbin2015,Erbin2021}
\begin{equation}
\nabla^{\nu}b_{\alpha\nu}=0, \quad \nabla^{\nu}c^{\alpha}+\nabla^{\alpha}c^{\nu}=0.
\end{equation}
Importantly, the ghost action vanishes on shell, so it only impacts the vertices of internal loop diagrams and does not affect the $S$-matrix.

Lastly, it is worth discussing the implications of associating quantum fluctuations with gauge transformations. Normally, fields related by gauge transformations are regarded as physically equivalent, and thus considered redundant in the path integral. To resolve this, one fixes the gauge and limits the integration to gauge-inequivalent fields. In the proposed approach, the situation is reversed: only gauge-equivalent metrics are included. This raises the question of why only gauge-equivalent fields should be counted for gravitation. The answer is to preserve unitarity and ensure background independence. In a renormalizable theory with a dimensionless coupling constant, the quantum corrections leave the form of the action invariant, and only unitary divergences appear. In gravity, the dimensional coupling constant leads inevitably to nonunitary higher derivative terms. Thus, the only way to preserve unitarity in quantum gravity is to eliminate the divergences altogether by limiting the quantum corrections to gauge transformations. Moreover, this parameterization renders the effective action independent of the on-shell condition for the background fields, ensuring the fundamental property of background independence is maintained. 

\section{Summary}
Within the background field method of quantum gravity, I have shown that if the quantum fluctuations are limited to diffeomorphic gauge transformations rather than the physical degrees of freedom, as in conventional quantum field theory, the quantum corrections to the effective action reduce to a delta function in the on shell condition. Therefore, on-shell the effective action is equal to the classical action with no quantum corrections. The use of such an unconventional prescription in quantum gravity is motivated by the unresolved conflict between renormalizability and unitarity. When coupled to matter with a conserved stress tensor, I show that an additional reparameterization ghost must be included for consistency. In addition, I have shown that this parameterization renders the effective action independent of the on shell condition for the background field, which is a form of background independence. Thus, a clear connection is established between background independence and unitarity and renormalizability in quantum gravity.

\bibliography{ms.bbl}

\end{document}